# Fourier temporal ghost imaging

**HONGHAO HUANG,**[1,2,†] **CHENGYANG HU,**[1,2,†] **MINGHUA CHEN,**[1,2] **SIGANG YANG,**[1,2] **AND HONGWEI CHEN**[1,2,*]

[1] *Department of Electronic Engineering, Tsinghua University, Beijing 100084, China*
[2] *Beijing National Research Center for Information Science and Technology (BNRist), Beijing 100084, China*
[†] *These authors contributed equally to this Letter.*

*\*chenhw@tsinghua.edu.cn*

**Abstract:** Ghost imaging is a fascinating framework which constructs the image of an object by correlating measurements between received beams and reference beams, none of which carries the structure information of the object independently. Recently, by taking into account space-time duality in optics, computational temporal ghost imaging has attracted attentions. Here, we propose a novel Fourier temporal ghost imaging (FTGI) scheme to achieve single-shot non-reproducible temporal signals. By sinusoidal coded modulation, ghost images are obtained and recovered by applying Fourier transformation. For demonstration, non-repeating events are detected with single-shot exposure architecture. It's shown in results that the peak signal-to-noise ratio (PSNR) of FTGI is significantly better (13dB increase) than traditional temporal ghost imaging in the same condition. In addition, by using the obvious physical meaning of Fourier spectrum, we show some potential applications of FTGI, such as frequency division multiplexing demodulation in the visible light communications.

## 1. Introduction

Ghost imaging (GI) is one of indirect measurement methods, which realizes spatial-resolved measurement via a detector without spatial resolving, therefore provides a way for reconstructing a signal beyond bandwidth limitation of the detector [1–8]. In a typical GI setup, imaging constructs an image by correlating two signals: one from the detection arm, which interacts with an object but possesses no spatial information (use bucket detector), and the other from the reference arm, which contains spatial information as probing patterns but does not interact with the object (use a space-resolving detector). The probing patterns may be random, in which case, they must be measured in a reference arm, or they can be pre-programmed prior to illuminating the object [9]. Recently, by taking into account space–time duality in optics, the concept of GI has been extended from space domain to time domain, which termed temporal ghost imaging (TGI) [10]. Whereafter, computational temporal ghost imaging (CTGI) with pre-modulate probing signals has been experimentally demonstrated. In CTGI, a fast photodiode is not required to record the time fluctuation of the light source, so the reference arm can be omitted. In [9,10], measurements over several thousand copies of the same temporal signal were necessary to retrieve a signal, limiting the applications only to the detection of reproducible signals. To detect non-repeating events, space-multiplexing [11] and wavelength-multiplexing [12] methods have been proposed and implemented to achieve single-shot TGI. The above methods can improve the universality of TGI, but poor signal-to-noise ratio (SNR) characteristics are an important issue for TGI. The methods of magnified TGI [13] and differential TGI [14] have been developed to circumvent SNR limitations, respectively. Although the above strategies can help a lot, they do not change the nature of TGI based on statistical model, which is determined by the use of random probing signals for illumination. The random probing signals form an over-complete non-orthogonal set, which results in a long-data acquisition time and low reconstruction quality. Even if the above-mentioned magnified or differential strategy is used, the SNR cannot be comparable with a conventional detector. An effective way to improve the reconstruction quality is to change the random probing signals to a deterministic probing signals, which has complete orthogonality. Among the strategies of deterministic models, Fourier ghost imaging (FGI) is a representative one, and its high reconstruction quality has been confirmed in spatial domain GI. [15].

In this paper, an attempt is made to implement the FGI in time domain, called Fourier temporal ghost imaging (FTGI). FTGI acquires temporal Fourier spectrum by using phase-shifting sinusoidal probing signals, and applies simple inverse fast Fourier transform algorithm to get the desired signal. In the experiment, we used the single-shot exposure architecture to obtain the spectrum of the temporal signal in one exposure which is performed by space-multiplexing of temporal signal. FTGI can detect non-repeating events through this strategy. Experimental results show that the quality of the reconstructed signal of FTGI is significantly better than that of traditional TGI, and the PSNR is increased by 13dB. In addition to demonstrating the high SNR advantage of FTGI, we also used the clear physical meaning of the Fourier spectrum to construct a frequency division multiplexing (FDM) decoding device, showing the potential of FTGI in the field of visible light communications.

## 2. Basic principle

The principle of the proposed FTGI technique is to obtain the spectrum of a time object and reconstruct the ghost image of it by performing inverse Fourier transform. To obtain the spectrum, the scene is encoded with pre-modulate phase-shifting sinusoidal probing signals and a bucket detector is used to collect the integrated light. The detected value $D_{k\varphi}$ of one measurement is equivalent to an Hadamard product of the time object $I(t)$ and the probing signal $S_{k\varphi}(t)$:

$$D_{k\varphi} = \langle I(t), S_{k\varphi}(t) \rangle = \int_{\tau} I(t)[A + B\cos(2\pi f_k t) + \varphi]dt \quad (1)$$

Where $S_{k\varphi}(t)$ is the sinusoidal probing signal with frequency $f$ and phase $\varphi$ and $\tau$ is the duration of the time object. $A$ and $B$ denote the average intensity and the contrast of $S_{k\varphi}(t)$, respectively. The Fourier coefficient $F_k$ of $f_k$ can be extracted by 4-step phase-shifting as

$$2BC \times F_k = (D_{k0} - D_{k\pi}) + i(D_{k\frac{\pi}{2}} - D_{k\frac{3\pi}{2}}) \quad (2)$$

Where $C$ depends on the response of the detector. The DC term $A$ can be cancelled out simultaneously by the 4-step phase-shifting. To form the spectrum of the time object, multiple Fourier coefficients are required to be assembled by using the above method and then be combined as:

$$F = \{F_n^*, F_{n-1}^*, ..., F_{n-1}, F_n\} \quad (3)$$

Where $n$ is the number of frequencies and $F_n^*$ denotes the complex conjugate of $F_n$. The time object $I(t)$ can be reconstructed by applying inverse Fourier transform:

$$2BC \times R = \mathfrak{I}^{-1}\{F\} \quad (4)$$

Where $\mathfrak{I}^{-1}$ denotes the inverse Fourier transform operator. The result of the inverse transform $R$ is proportional to the time object $I(t)$.

To acquire sufficient $n$ Fourier coefficients to form the spectrum, measurements over multiple copies of the same temporal object is needed, which can be realized by both serial and single-shot (parallel) CTGI schemes. In serial CTGI scheme, different patterns modulate the time object successively in time domain to make measurements, which requires the time object to be reproducible and synchronized. We build the FTGI in a single-shot scheme, that is, we utilize a long-exposure image sensor to perform space-multiplexing parallel measurements that allow the acquisition of a nonreproducible time object.

## 3. Experiments and discussion

### 3.1 Experimental setup

The experimental setup is shown in Fig. 1(a). A light-emitting diode (LED, Thorlabsm MCWHL6-C1) is driven by a time-varying signal generated by an arbitrary waveform generator (AWG, UNI-T UTG2000B) and its emitted light intensity is linearly modulated as a time object. The modulated light illuminates a digital micro mirror device (DMD, ViALUX V-9001) and encoded by a set of pre-designed patterns. The reflected light from DMD is then focused onto an image sensor (FLIR GS3-U3-120S6M-C) by a zoom lens (Utron VTL0714V). As shown in Fig. 1(b), to realize parallel phase-shifting sinusoidal probing, a series of patterns is computationally generated. The grayscales of these patches vary in discrete sinusoidal fashion with a sampling rate of 800 patterns per second. The patterns are spatially divided into 10 x 10 patches corresponding to 100 Fourier coefficients acquisition and the corresponding frequencies of these 100 patches are set from direct current (0 Hz) to 99 Hz with 1 Hz step. Each patch consists of 2 x 2 sub-patches corresponding to 4 phases (0, 0.5pi, pi, 1.5pi). Therefore, there are 20 x 20 sub-patches in total, each of which serves as a measurement. For the direct current patches, the grayscales are the sinusoidal value of these 4 phases. Although the DMD only modulates the light in a binary form, we utilize its pulse-width modulation (PWM) mode to realize equivalent temporal sinusoid modulation with the generated grayscale patterns. Since a single exposure of the image sensor integrates the Hadamard product between sinusoid signal and temporal waveform of the scene, we can extract the Fourier coefficient for one specific frequency by means of 4-step phase-shifting as equation (2) and then form the spectrum to reconstruct the ghost image of the time object.

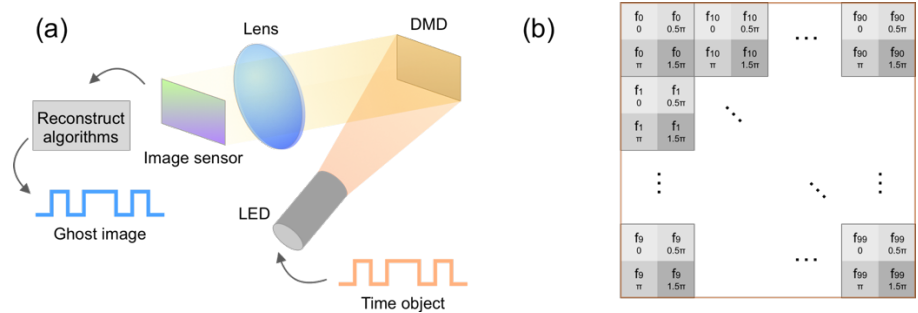

Fig. 1. Overview of FTGI. (a) Experimental setup. DMD: digital micro mirror device. LED: light-emitting diode. (b) An illustration of DMD encoding pattern with the corresponding frequencies and phases marked in each patch.

Basic sinusoidal waveforms with one second duration and different frequencies (11, 33, 55, 88 Hz) are used as time objects to validate the system and demonstrate the principle of FTGI. The raw captures from the image sensor are shown in Fig. 2(a). In each raw capture, the top-left patch (red) shows the direct current. The patch green box denotes the patch with the same frequency to the time object in each capture. The spectrums obtained from the raw captures are shown in Fig. 2(b). Except to the patches denoted by the red and green boxes, all patches consist of four sub-patches with the similar grayscale value, resulting to these Fourier coefficients being near to zero, which is consistent of the single-frequency nature of sinusoids. The peaks in the spectrums clearly denote the frequency of the sinusoidal waveforms. By applying inverse Fourier transform, the ghost images of the time objects are reconstructed as shown in Fig. 2(c).

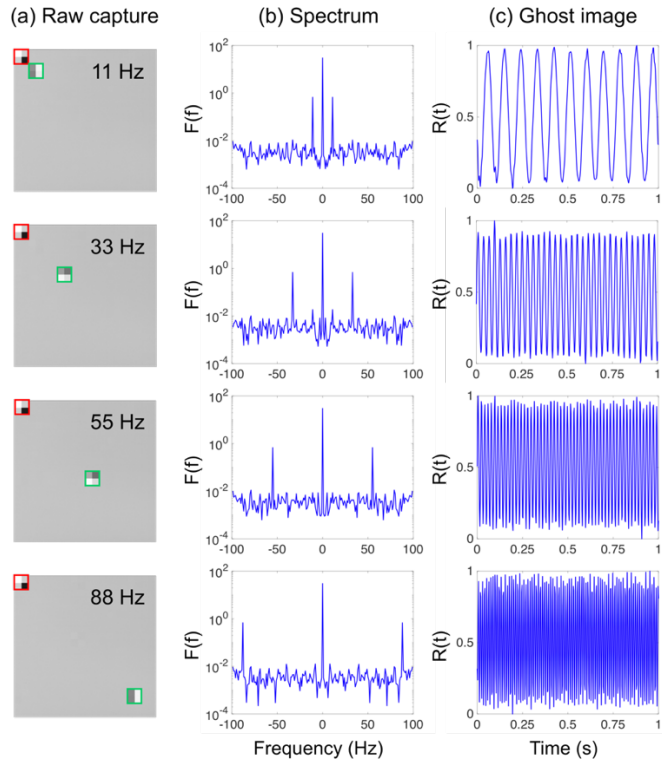

Fig. 2. Using FTGI to capture sinusoidal time objects with frequencies of 11, 33, 55, 88 Hz. (a) Raw capture data from image sensor. Red boxes: direct current patch. Green boxes: patches corresponding to the frequencies of the time objects. (b) Extracted spectrum. (c) Reconstructed ghost image.

### 3.2 Basic waveforms results

Three kinds of basic periodic waveforms, i.e., square wave, sawtooth wave and pulse wave (with 10% duty cycle) with frequencies of 2, 5, 7, 11 Hz with one second duration are used as time objects in experiments. Ghost images reconstructed by FTGI (blue lines) as well as ground truth references (red dash lines) are shown in Fig. 3. The whole raw capture data from the image sensor is used to reconstruct the ghost images thus the bandwidth is 99 Hz. Because all Fourier coefficients share the same weights in the demonstration, which deploys an equivalent rectangle window filter to the system, ringing effect exists in the rising edges and the falling edges of the reconstructed waveforms, which can be suppressed by applying roll-off attenuation to the Fourier coefficients.

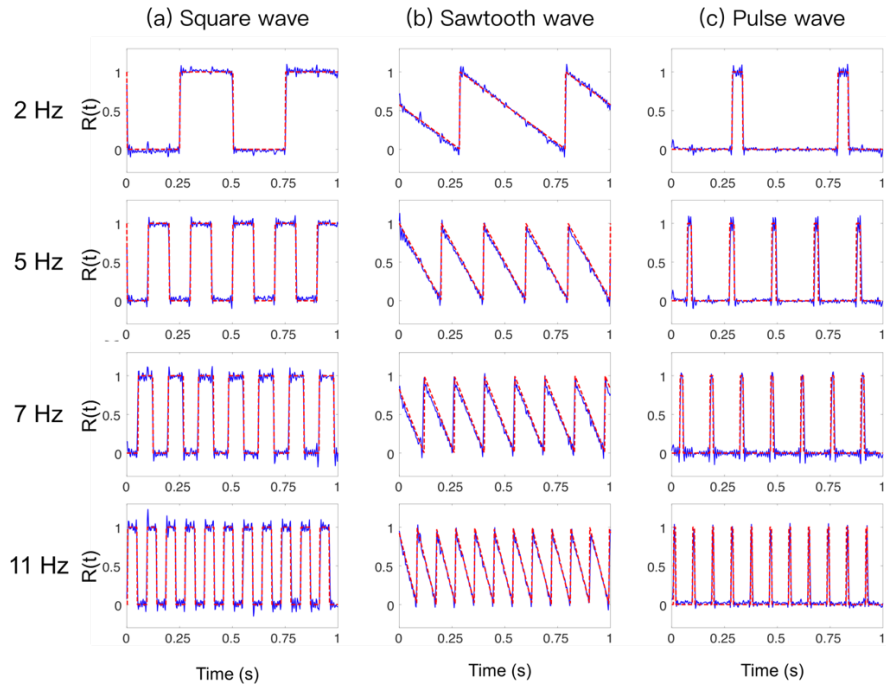

Fig. 3. Reconstructed ghost images of basic function waveforms. (a) Square wave. (b) Sawtooth wave. (c) Pulse wave. The frequencies of the signals are 2, 5, 7, 11 Hz from top to bottom, respectively. Blue solid lines: reconstructed ghost images. Red dash lines: ground truth references.

The proposed FTGI system, acquiring the Fourier spectrum in the order from lower to higher frequencies, is a compressive sampling like approach. Similar to the compressive ratio in compressive ghost imaging, we define a spectrum coverage for FTGI. For the demonstration in this paper, a 100% spectrum coverage corresponds to a bandwidth of 99Hz. Since the spectrum is obtained by a single shot of the imaging sensor, one can change the spectrum coverage to change the data size by adjust the imaging region of the sensor. This insight can help us find a balance between data size and the number of measurements when dealing with different application scenarios. Fig. 4(a)-(c) show the reconstructed ghost imaging of three kinds of waveform with different spectrum coverages from 25% to 100%. With a low spectrum coverage as 25%, the outline of the time objects can be recognized. As more spectrum components are covered, more details of the time objects can be maintained to improve the reconstruction quality.

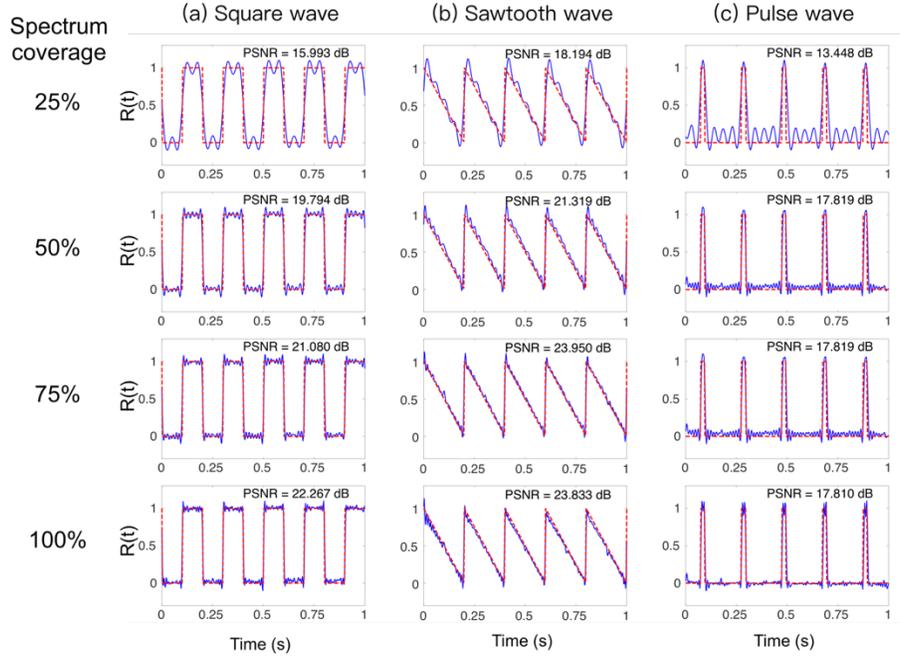

Fig. 4. Reconstructed ghost images with different spectrum coverage. The frequency of time objects is 5 Hz and the spectrum coverages are from 25% to 100% (top to bottom in the figure). Blue solid lines: reconstructed ghost images. Red dash line: ground truth references. (a) Square wave. (b) Sawtooth wave. (c) Pulse wave.

The peak signal-to-noise ratio (PSNR) is introduced to quantitatively evaluate the reconstruction, which is defined as:

$$PSNR = 10\log_{10}\left[\frac{P^2}{\frac{1}{L}\sum_i (R_i - G_i)}\right] \quad (5)$$

Where $R$ and $G$ are reconstructed ghost image and ground-truth reference, respectively. $P$ and $G$ is the maximum of $R$. $L$ is the length of $R$ and $G$. A larger PSNR means the ghost image is closer to the original time object.

In Fig. 5(a)-(c), the PSNR values of three kinds of waveform are calculated for different spectrum coverages and signal frequencies. For all curves in the figure, the general trend of the PSNR appears to be increasing with the spectrum coverage. In details, the PSNR grows up rapidly with the spectrum coverage increasing from 0% to around 40% and it tends to be stable when the spectrum coverage is over around 50%. In Fig. 5(a), the curves corresponding to four frequencies of square wave present ladder form rises, for the reason that the spectrum of a periodic square signal is discrete and the spectrum components between the harmonics does not contains additional information of the signal. Furthermore, under the same spectrum coverage, the lower time object frequency leads to a higher PSNR due to the narrower bandwidth it has. For sawtooth wave (Fig. 5(b)) and pulse wave (Fig. 5(c)), they have similar trends to the square wave on the whole.

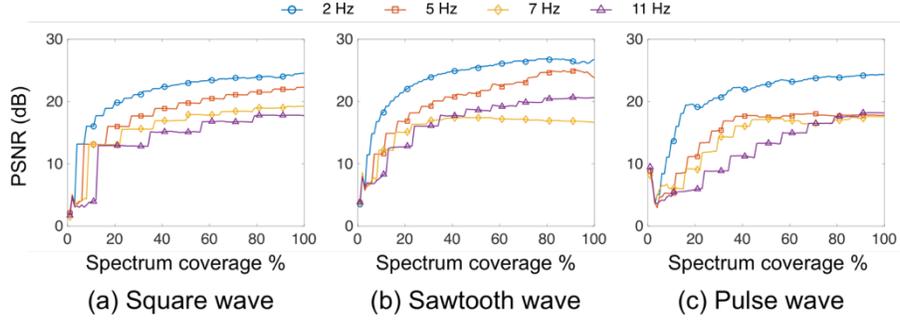

Fig. 5. PSNR of square, sawtooth and pulse waves varying with spectrum coverage. (a) Square wave. (b) Sawtooth wave. (c) Pulse wave.

We also demonstrate the ability of FTGI to reconstruct ghost image of aperiodic time objects. The time object is a 10-bit binary word "1110010110" with one second duration. The binary word is encoded as non-return-to-zero (NRZ) square code (Fig. 6(a)), return-to-zero (RZ) square code (Fig. 6(b)) and return-to-zero (RZ) Gauss code (Fig. 6(c)), respectively. The corresponding bit sequence is shown on the top of each plot. The ghost images (blue lines) well fit the ground truth references (red dash lines) with a high PSNR. The PSNR of RZ Gauss code is higher than square codes thanks to its roll-off nature in Fourier domain which eases the ringing effect.

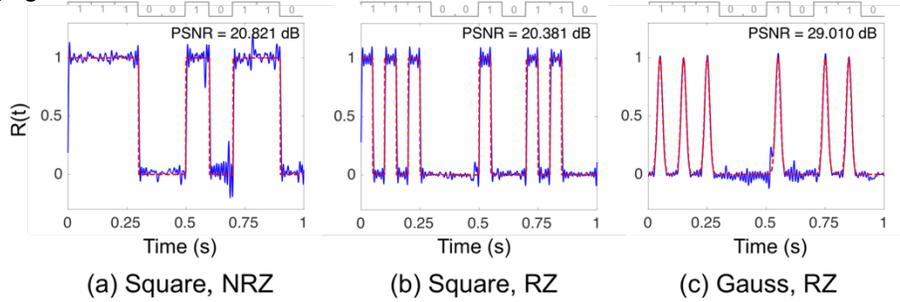

Fig. 6. Reconstructed ghost images of a binary word of 10 bits. The bit sequence used to generate the signals is shown on the top. NRZ: non-return-to-zero. RZ: return-to-zero. Blue lines: reconstructed ghost images. Red dash line: ground truth references. (a) NRZ square code. (b) RZ square code. (c) RZ Gauss code.

*3.3 Comparison with random probing TGI*

Experiments are performed to compare FTGI to random-probing TGI. To make a fair comparison, all setup including devices and number of measurements (i.e., 20 x 20, 400 measurements in total) of FTGI and TGI are kept the same except DMD encoding patterns and reconstruction algorithms. Each measurement stands for one detected value $D_{k\varphi}$ with specific frequency and phase in FTGI, or one realization in TGI. Thus, the measurement percentage corresponds to spectrum coverage in FTGI or comprehensive ratio in TGI. For the TGI experiment, independent random binary patterns are loaded to the DMD to encode the time objects and intensity correlation method used in [10] is deployed to reconstruct the ghost images. The RZ Gauss encoded 10-bit binary word "1110010110" is used as an exemplar time object.

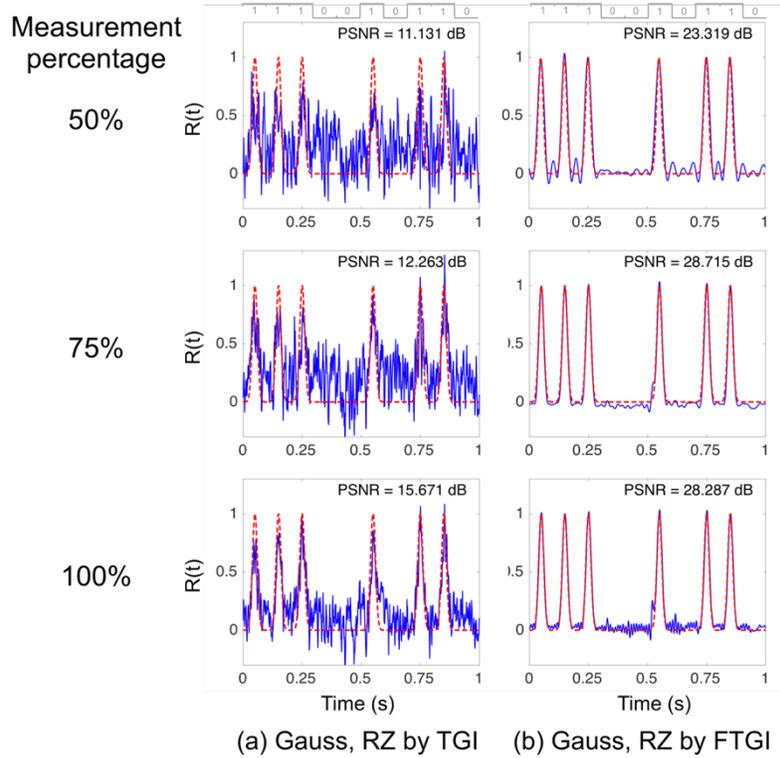

Fig. 7. Comparison between TGI and FTGI results. Blue solid lines: reconstructed ghost images. Red dash line: ground truth references. (a) RZ Gauss code ghost image by TGI. (b) RZ Gauss code ghost image by FTGI.

Results reconstructed by TGI is shown in Fig. 6(a). When the percentage of measurements is relatively low (50%), the waveform suffers from much noise, where severe spikes and glitches will cause error of representing the binary information. With the more measurements being taken, the PSNR gets better and the peaks corresponding to "1" in the binary word becomes clearer. However, burrs still exist on the edge of the peaks. By contrast, in the respectively of FTGI, the PSNR is much higher than that of TGI in the same condition and the reconstructed ghost images clearly and precisely depict the time objects. The peaks in the FTGI results are smoother than those in TGI results. When the measurement percentage increases from 75% to 100%, the PSNR slightly decreases because more noise introduced with larger bandwidth filters used. Still, these two results show high PSNR over 28 dB. These results demonstrate, FTGI enables a higher quality time objects reconstruction than TGI in the same condition.

*3.4 FDM decoding demonstration*

Here we also demonstrate using FTGI to decode FDM signals, as one of the application scenarios of this technique. Two binary words are modulated onto two carrier frequencies, i.e. 25 Hz for channel 1 and 75 Hz for channel 2, and combined together to be an FDM signal. With the same experimental setup and the same protocol, the FDM signal drives the LED to generate light signal and further detected by FTGI. As shows in Fig. 8(a), to decode the FDM signal, the imaging sensor capture is divided into two parts for channel 1 and channel 2, respectively. The whole spectrum extracted from the capture (shown in Fig. 8(b) and Fig. 8(c), for 5 bit/s and 10 bit/s, respectively) is split to form the spectrums for the two channels. By performing inverse Fourier transform for each spectrum, signals transmitted in each channel can be decoded.

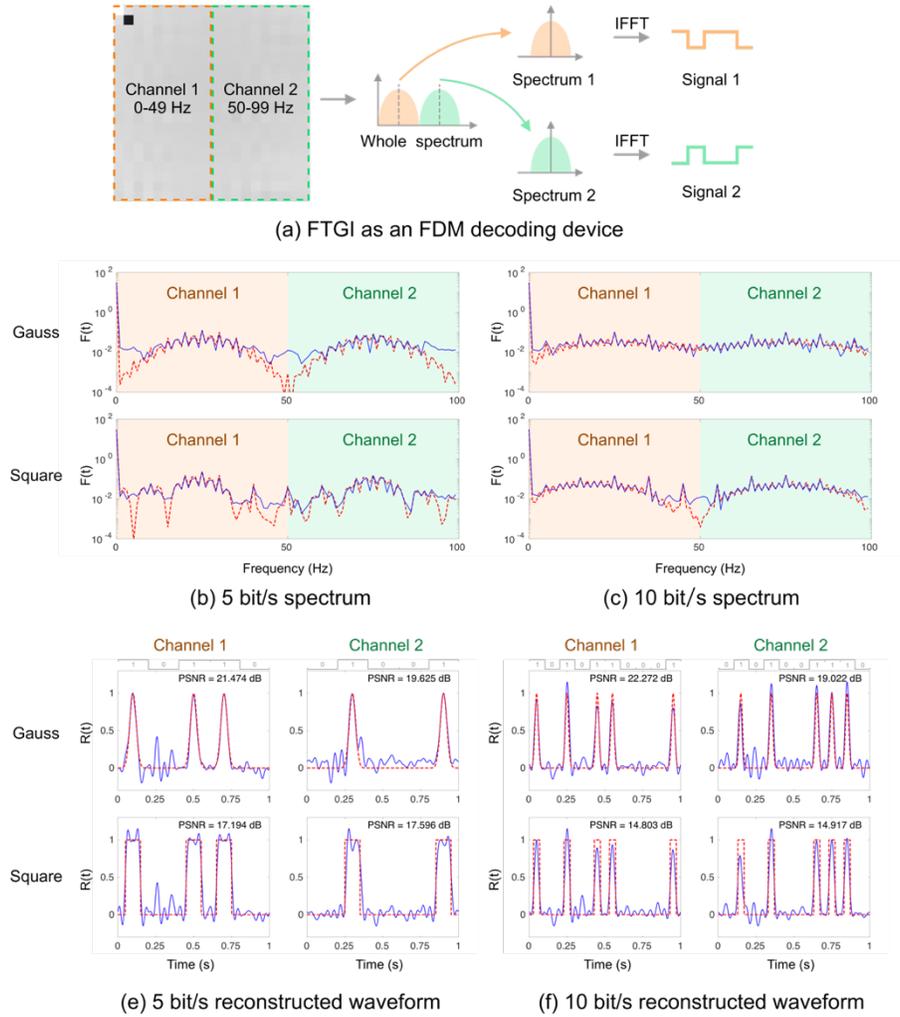

Fig. 8. Demonstration of FDM decoding. (a) Illustration of FDM decoding. (b) Spectrums of 5 bit/s gauss and square RZ signals. (c) Spectrums of 10 bit/s gauss and square RZ signals. (d) Reconstructed waveforms of 5 bit/s gauss and square RZ signals. (f) Reconstructed waveforms of 10 bit/s gauss and square RZ signals. Blue solid lines: reconstructed ghost images. Red dash line: ground truth references.

Two bit-rate setups are used to test the system, that is, 5 bit/s (Fig. 8(e)) and 10 bit/s (Fig. 8(f)) for each channel. For both bit-rate, the PSNR of Gauss code is higher than that of square code. When the bit-rate improved from 5 bit/s to 10 bit/s, the PSNR of Gauss code keeps stable but that of square code decreases by around 3 dB. Nevertheless, instead of introducing more noise, the reduction of PSNR of square code is caused by the distortion of the waveform due to the limited bandwidth, which deforms the waveform from standard square wave. Therefore, the binary word can still be recognized although the PSNR is relatively low.

## 4. Conclusion

To summarize, these experiments represent the best SNR demonstration, to the best of our knowledge, of TGI. They are performed by acquiring the Fourier spectrum of the desired temporal signal with the use of phase-shifting sinusoidal probing signals. Besides the repeatable signal applications, FTGI is used for single-shot acquisition of non-reproducible temporal signal, based on spatial multiplexed measurement. We demonstrate that single-shot FTGI can also be used as a frequency division multiplexing decoding device to separate and reconstruct accurately frequency-multiplexed temporal signals. In terms of computational efficiency, the signal reconstruction employs the use of one-dimensional (1D) inverse fast Fourier transform (IFFT) algorithm rather than any iterative or minimization algorithms. Due to low computational complexity, IFFT can be used to reconstruct signal from a single exposure in real-time.

Beyond the high-quality temporal ghost imaging demonstrated above, the proposed FTGI can actually be extended to be a flexible Fourier domain sampling framework. By adjusting weights to the probing signals of different frequencies, filters can be deployed to the process of signal acquisition to get filtered results directly. What is more, one can precisely choose the frequencies of interest to be obtained instead of equidistant frequency distribution, which gives the potential to build a smarter signal sensing system.

The temporal resolution is a key parameter of TGI, which is determined by the bandwidth of the modulator. In the present form of the device (i.e., single-shot FTGI), its obvious feature is the slowness caused by the equipment (i.e., DMD) used to produce the sinusoidal mode. To overcome this drawback, the time-encoded methods [16,17] consisting of time-stretch and inertia-free spectrally scanning can be used to improve detection frequency band. Although the 4-step phase-shifting offers better measurement performance while one can also utilize 3-step phase-shifting [18] or 2-step phase-shifting [19] for a higher detection efficiency. In addition to frequency division multiplexing decoding, features such as low-pass filtering, high-pass filtering, and band-pass filtering can also be implemented in FTGI. As a framework for directly collecting spectrum, FTGI has potential in the field of optical communication and structured detection [20,21].

**Disclosures**

The authors declare no conflicts of interest.